\documentclass[acmsmall,screen,nonacm]{acmart}
\usepackage{pgfplots}
\usepackage{pgfplotstable}
\pgfplotsset{compat=1.18}
\usepackage{xcolor}
\usepackage{listings}

\definecolor{tan}{HTML}{BCAAA4}
\definecolor{lava}{HTML}{EF6C00}
\definecolor{codekeyword}{HTML}{0B5394}
\definecolor{codetype}{HTML}{0E6B6E}
\definecolor{codecomment}{HTML}{3D7A3A}
\definecolor{codestring}{HTML}{7A1E76}
\definecolor{codeoperator}{HTML}{4A5568}
\definecolor{codefunction}{HTML}{8A4B08}
\definecolor{codeargument}{HTML}{1F4E79}

\lstdefinelanguage{Rocq}{
  sensitive=true,
  morekeywords={
      Definition,Conjecture,QuickChick,forall,if,then,else
    },
  morecomment=[s]{(*}{*)},
  morestring=[b]",
  literate=
      {:=}{{{\color{codeoperator}\bfseries:=}}}2
      {=>}{{{\color{codeoperator}\bfseries=>}}}2
      {->}{{{\color{codeoperator}\bfseries->}}}2
      {<-}{{{\color{codeoperator}\bfseries<-}}}2
      {|}{{{\color{codeoperator}\bfseries|}}}1
}

\lstdefinestyle{thesiscode}{
  basicstyle=\ttfamily\small\linespread{1.06}\selectfont,
  aboveskip=\baselineskip,
  belowskip=\baselineskip,
  keywordstyle=\color{codekeyword}\bfseries,
  keywordstyle=[2]\color{codetype}\bfseries,
  identifierstyle=\color{black},
  emphstyle=\color{codefunction}\bfseries,
  emphstyle=[2]\color{codeargument}\itshape,
  commentstyle=\color{codecomment}\itshape,
  stringstyle=\color{codestring},
  showstringspaces=false,
  keepspaces=true,
  columns=fullflexible
}

\newcommand{\codeinline}[2][language=Rust]{\lstinline[style=thesiscode,#1]`#2`}
\lstnewenvironment{code}[1][language=Rust]{\lstset{style=thesiscode,#1}}{}

\makeatletter
\let\code@envbegin\code
\newcommand{\code@inline}{%
  \@ifnextchar[{\code@inline@opts}{\code@inline@opts[language=Rust]}%
}
\def\code@inline@opts[#1]#2{%
  \codeinline[#1]{#2}%
}
\DeclareRobustCommand{\code}{%
  \def\code@envname{code}%
  \ifx\@currenvir\code@envname
    \let\code@next\code@envbegin
  \else
    \let\code@next\code@inline
  \fi
  \code@next
}
\makeatother

\AtBeginDocument{%
  }

\begin{document}

\title{QuickerChick}

\author{Ivan Mladenov}
\orcid{0009-0006-6132-6700}
\affiliation{%
  \institution{University of Maryland}
  \city{College Park}
  \state{MD}
  \country{USA}
}

\author{Alperen Keles}
\orcid{0009-0000-5734-3598}
\affiliation{%
 \institution{University of Maryland}
 \city{College Park}
 \state{MD}
 \country{USA}}

\author{Leonidas Lampropoulos}
\orcid{0000-0003-0269-9815}
\affiliation{%
  \institution{University of Maryland}
  \city{College Park}
  \state{MD}
  \country{USA}}

\renewcommand{\shortauthors}{Mladenov et al.}

\begin{abstract}
    Property-based testing (PBT) with QuickChick relies 
    on extracting Rocq programs to OCaml. However, this
    extraction mechanism, while crucial for QuickChick
    to function, has significant performance implications.
    In this work, we describe how we optimized QuickChick, exploiting various opportunities 
    offered by program extraction to significantly improve 
    each test's extraction, compilation, and running time. 
    We also evaluate these improvements using the ETNA 
    benchmarking platform for PBT to assess how individual
    improvements impacted the overall test run time. 
\end{abstract}




\maketitle

\section{Introduction}
{\itshape QuickChick} \cite{Lampropoulos18} is the canonical library for property-based testing in Rocq. A programmer 
writes a property that they expect should hold, and QuickChick 
generates test cases to verify said property. Normally, a PBT 
library will run these tests in the language in which the tests 
were written --- but not QuickChick! That's what makes QuickChick unique. 

QuickChick needs to use a mechanism called {\itshape extraction} 
\cite{Letouzey02} to be able to execute tests. Rocq, as a 
proof assistant, lacks some of the side-effecting capabilities 
that are necessary for random testing (such as randomness or IO). 
Extraction allows QuickChick to convert a test written in Rocq 
directly into (hopefully) equivalent OCaml code, which can then 
be compiled and executed with its results reported back to 
the user within an interactive proving session. For example, consider
the following simple property, which checks that the result of a \texttt{max}
function is greater than both of its inputs:

\begin{code}[language=Rocq]
    Definition max (x : nat) (y : nat) : nat := if x <=? y then y else x.
    
    Conjecture prop_maxGe: forall (x y : nat), max x y >= y /\ max x y >= x.
    
    QuickChick prop_maxGe.
\end{code}
 When a user runs the QuickChick command on the last line, 
they will get some output like this:
\begin{code}[language=Rocq]
    > QuickChecking prop_maxGe
    > 
    > +++ Passed 10000 tests (0 discards)
\end{code}

Under the hood, this property is extracted to a monolithic 
OCaml file, which is then compiled to run the tests. Let's 
look at the size of said file:
\begin{code}[language=Rocq]
    $ wc -l prop_maxGe.ml
        2760 prop_maxGe.ml
\end{code}

How does a single property end up generating such a large file?
For the OCaml program to run, every single dependency---whether
from QuickChick itself, the standard library, or the core library---of the property must also be extracted to the file. This introduces 
some obvious overhead. For one, the extraction scheme must recursively
walk all library dependencies, making the extraction time longer. On
top of that, every extracted dependency must be compiled to execute
tests, i.e., building these large singleton files wastes time.

There's a silver lining to all of this: many of these dependencies
are frequently reused functions and type definitions, which means
one could pre-compile them into a library and link against it. This
is precisely what we have done.

\section{An Extracted Library}
We developed an internal library for QuickChick, which contains 
OCaml equivalents to many of QuickChick's commonly used dependencies.
Having these library definitions gave us many opportunities to
squeeze out better performance.

The extraction library in Rocq allows users to hijack extraction by
replacing terms or types directly with raw strings of OCaml code.
This lets us replace any dependencies with 
direct references to our library definitions, ultimately shortening
the recursive walk. With the example from before, we
can see how much the file shrank by inlining our library functions:
\begin{code}[language=Rocq]
    $ wc -l prop_maxGe.ml
         218 prop_maxGe.ml
\end{code}

From our experiments, we saw a 1.5-3$\times$ improvement in extraction,
along with a roughly 1.2$\times$ improvement in compile time. So despite 
the drastic reduction in file size, the compile time didn't reflect these
changes as much as we'd hoped. As it turns out, the OCaml compiler is 
incredibly effective; rather, the slowdown was in the call to {\itshape ocamlbuild}.
We thus changed the build to directly compile with {\itshape ocamlopt}
and explicitly link against the library, resulting in a >2$\times$ improvement
in compile time. We can see the improvements already through
ETNA \cite{Keles23} experiments for binary search trees (BST), red-black trees (RBT), and
the simply-typed lambda calculus (STLC):

\begin{figure*}[!ht]
\centering

\begin{tikzpicture}[baseline=-33.5pt]
\begin{axis}[
    title={\small BST},
    ybar,
    bar width=10pt,
    width=3.75cm,
    height=3.75cm,
    symbolic x coords={Extract, Compile},
    xtick=data,
    xticklabel style={font=\scriptsize},
    yticklabel style={font=\scriptsize},
    ylabel={Average Time (ms)},
    ylabel style={font=\scriptsize},
    ymin=0,
    legend style={
        font=\scriptsize,
        at={(0.5,-0.25)},
        anchor=north,
        legend columns=2,
    },
    enlarge x limits=0.4,
    grid=major,
    grid style={line width=0.3pt, draw=gray!20},
]
\addplot[fill=tan] coordinates {
    (Extract, 132)
    (Compile, 667)
};
\addplot[fill=lava] coordinates {
    (Extract, 92)
    (Compile, 303)
};
\end{axis}
\end{tikzpicture}
\hfill
\begin{tikzpicture}
\begin{axis}[
    title={\small RBT},
    ybar,
    bar width=10pt,
    width=3.75cm,
    height=3.75cm,
    symbolic x coords={Extract, Compile},
    xtick=data,
    xticklabel style={font=\scriptsize},
    yticklabel style={font=\scriptsize},
    ymin=0,
    legend style={
        font=\scriptsize,
        at={(0.5,-0.30)},
        anchor=north,
        legend columns=2,
    },
    enlarge x limits=0.4,
    grid=major,
    grid style={line width=0.3pt, draw=gray!20},
]
\addplot[fill=tan] coordinates {
    (Extract, 146)
    (Compile, 726)
};
\addplot[fill=lava] coordinates {
    (Extract, 52)
    (Compile, 366)
};
\legend{Original, Optimized}
\end{axis}
\end{tikzpicture}
\hfill
\begin{tikzpicture}[baseline=-33.5pt]
\begin{axis}[
    title={\small STLC},
    ybar,
    bar width=10pt,
    width=3.75cm,
    height=3.75cm,
    symbolic x coords={Extract, Compile},
    xtick=data,
    xticklabel style={font=\scriptsize},
    yticklabel style={font=\scriptsize},
    ymin=0,
    legend style={
        font=\scriptsize,
        at={(0.5,-0.25)},
        anchor=north,
        legend columns=2,
    },
    enlarge x limits=0.4,
    grid=major,
    grid style={line width=0.3pt, draw=gray!20},
]
\addplot[fill=tan] coordinates {
    (Extract, 127)
    (Compile, 690)
};
\addplot[fill=lava] coordinates {
    (Extract, 56)
    (Compile, 342)
};
\end{axis}
\end{tikzpicture}

\caption{Average extraction and compilation times (ms) for BST, RBT, and STLC benchmarks (ETNA)}
\label{fig:fixed-costs}
\end{figure*}
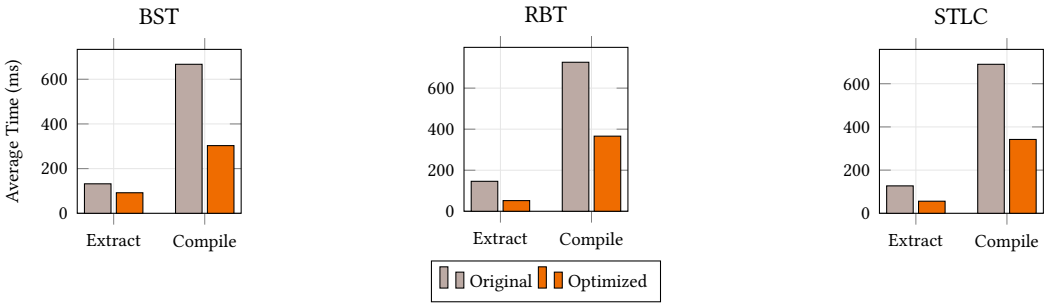

Writing this library also gave rise to an opportunity not previously
available: tail-call optimizations. Now that we're writing in OCaml
rather than Rocq, we could rewrite naively recursive functions and
let the compiler optimize them out. This change alone resulted in 
1.2-1.8$\times$ improvement in run time. While looking for these 
inefficiencies, we realized that the function responsible for computing
test case sizes was too expensive for how frequently it's called.
Without significantly changing the distribution of test sizes,
we were able to get yet another 2$\times$ speedup. 

Lastly, it was worthwhile to experiment with different optimization
levels and compilers. QuickChick was already using the native compiler,
but we wanted to see how a compiler like {\itshape flambda} \cite{Flambda25} 
would perform. Evidently, the compilation time was slightly worse with the
aggressive optimizations. The question was whether the run-time benefits 
outweighed the compile-time penalty. The answer was a resounding yes.
Because we're already pre-compiling the library functions, and on top of
that we've cut down on the actual file size, the time spent compiling
one property test is negligible. With that, we get a free performance
boost in run time, another 1.5$\times$ speedup on average. 

We can zoom out now and see what all of these run-time improvements amount
to:

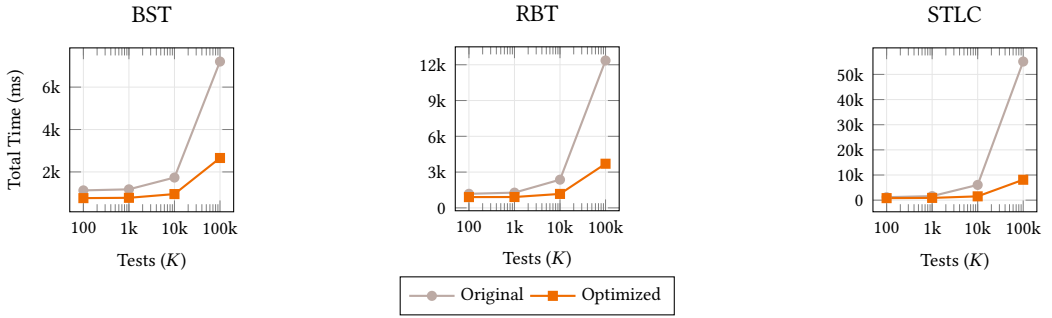
\begin{figure*}[!ht]
\centering
\begin{tikzpicture}[baseline=-39pt]
\begin{axis}[
    title={\small BST},
    width=3.75cm,
    height=3.75cm,
    xmode=log,
    ymode=normal,
    xlabel={Tests ($K$)},
    ylabel={Total Time (ms)},
    xlabel style={font=\scriptsize},
    ylabel style={font=\scriptsize},
    scaled y ticks=false,
    xticklabel style={font=\scriptsize},
    yticklabel style={font=\scriptsize},
    xtick={100, 1000, 10000, 100000},
    xticklabels={100, 1k, 10k, 100k},
    ytick={0, 2000, 4000, 6000},
    yticklabels={0, 2k, 4k, 6k},
    legend style={
        font=\scriptsize,
        at={(0.5,-0.35)},
        anchor=north,
        legend columns=2,
    },
    grid=major,
    grid style={line width=0.3pt, draw=gray!20},
    mark size=1.5pt,
]
\addplot[color=tan, mark=*, thick] coordinates {
    (100,    1129)
    (1000,   1179)
    (10000,  1735)
    (100000, 7206)
};
\addplot[color=lava, mark=square*, thick] coordinates {
    (100,    764)
    (1000,   779)
    (10000,  955)
    (100000, 2657)
};
\end{axis}
\end{tikzpicture}
\hfill
\begin{tikzpicture}
\begin{axis}[
    title={\small RBT},
    width=3.75cm,
    height=3.75cm,
    xmode=log,
    ymode=normal,
    xlabel={Tests ($K$)},
    xlabel style={font=\scriptsize},
    xticklabel style={font=\scriptsize},
    yticklabel style={font=\scriptsize},
    scaled y ticks=false,
    xtick={100, 1000, 10000, 100000},
    xticklabels={100, 1k, 10k, 100k},
    ytick={0, 3000, 6000, 9000, 12000},
    yticklabels={0, 3k, 6k, 9k, 12k},
    legend style={
        font=\scriptsize,
        at={(0.5,-0.40)},
        anchor=north,
        legend columns=2,
    },
    grid=major,
    grid style={line width=0.3pt, draw=gray!20},
    mark size=1.5pt,
]
\addplot[color=tan, mark=*, thick] coordinates {
    (100,    1180)
    (1000,   1277)
    (10000,  2364)
    (100000, 12356)
};
\addplot[color=lava, mark=square*, thick] coordinates {
    (100,    904)
    (1000,   910)
    (10000,  1170)
    (100000, 3701)
};
\legend{Original, Optimized}
\end{axis}
\end{tikzpicture}
\hfill
\begin{tikzpicture}[baseline=-39pt]
\begin{axis}[
    title={\small STLC},
    width=3.75cm,
    height=3.75cm,
    xmode=log,
    ymode=normal,
    xlabel={Tests ($K$)},
    xlabel style={font=\scriptsize},
    xticklabel style={font=\scriptsize},
    yticklabel style={font=\scriptsize},
    scaled y ticks=false,
    xtick={100, 1000, 10000, 100000},
    xticklabels={100, 1k, 10k, 100k},
    ytick={0, 10000, 20000, 30000, 40000, 50000},
    yticklabels={0, 10k, 20k, 30k, 40k, 50k},
    legend style={
        font=\scriptsize,
        at={(0.5,-0.35)},
        anchor=north,
        legend columns=2,
    },
    grid=major,
    grid style={line width=0.3pt, draw=gray!20},
    mark size=1.5pt,
]
\addplot[color=tan, mark=*, thick] coordinates {
    (100,    1177)
    (1000,   1610)
    (10000,  6062)
    (100000, 55118)
};
\addplot[color=lava, mark=square*, thick] coordinates {
    (100,    809)
    (1000,   898)
    (10000,  1515)
    (100000, 8117)
};
\end{axis}
\end{tikzpicture}
\caption{Average total runtime (ms) vs test counts for BST, RBT, and STLC benchmarks (ETNA), semi-log}
\label{fig:variable-costs}
\end{figure*}

With all of our optimizations in hand, we achieved a 3-7$\times$
overall speedup for QuickChick. In all benchmarks, even with small
test cases, the optimized library finished tests faster. No matter
the price paid for compilation time, the overall time improvements 
lead to a snappier testing experience. Where previously executing 100,000 tests
of certain properties could take close to a minute, with the combination 
of our optimizations, the same testing runs can finish in under 10 seconds.

\section{Conclusion and Future Work}
Ideally, we'd have a 1-1 equivalent of all QuickChick library functions 
written, hand-optimized, and compiled in OCaml. Because this library is 
extensible, we can keep translating Rocq functions to approach this goal, 
in the meantime improving performance and the user experience. Ultimately, 
this extracted library lays the groundwork for a better-performing QuickChick, 
where the gap between testing verified programs and efficient execution 
is narrowing.

\bibliographystyle{ACM-Reference-Format}
\bibliography{QuickerChick}

\appendix

\end{document}